\documentclass[aps,prl,amsmath,amssymb,showpacs,superscriptaddress,twocolumn]{revtex4-1}

\usepackage{verbatim,bbm,graphicx,color}

\newcommand{\bra}[1]{\ensuremath{\langle #1 |}}
\newcommand{\ket}[1]{\ensuremath{| #1 \rangle}}
\newcommand{\be}{\begin{equation}}
\newcommand{\ee}{\end{equation}}
\newcommand{\ie}{{\it i.e.}\ }
\newcommand{\eg}{{\it e.g.}\ }
\newcommand{\dd}{\ensuremath{\text{d}}}

\begin{document}

\title{Adaptive Resummation of Markovian Quantum Dynamics}
\author{Felix Lucas}
\affiliation{Max Planck Institute for the Physics of Complex Systems, N\" othnitzer Stra\ss{}e 38, 01187 Dresden, Germany}
\affiliation{University of Duisburg-Essen, Faculty of Physics, Lotharstra\ss{}e 1-21, 47057 Duisburg, Germany}
\author{Klaus Hornberger}
\affiliation{University of Duisburg-Essen, Faculty of Physics, Lotharstra\ss{}e 1-21, 47057 Duisburg, Germany}

\date{\today}

\begin{abstract}
We introduce a method for obtaining analytic approximations to the evolution of Markovian open quantum systems. It is based on resumming a generalized Dyson series in a way that ensures optimal convergence even in the absence of a small parameter. The power of this approach is demonstrated by two benchmark examples: the spatial detection of a free particle and the Landau-Zener problem in the presence of dephasing. The derived approximations are asymptotically exact and exhibit errors on the per mil level over the entire parameter range.
\end{abstract}
 
\maketitle

\emph{Introduction.---} Emergent quantum technologies employ ever larger and noisier quantum machines, such as solid state devices \cite{Lukin2006}, complex biomolecules \cite{Rabitz2009}, or high precision detectors \cite{Napolitano2011}. While it is hard to shield such systems from environmental influences, physicists have come to think of the latter as a potentially useful resource in areas ranging from quantum chemistry \cite{Prezhdo2000,Rabitz2006} to quantum optics \cite{Wiseman2001,Moelmer2007,Morigi2007} and quantum information \cite{Cirac2009}. 

If one seeks to exploit incoherent effects beyond the preparation of stationary states \cite{Wiseman2001,Viola2009,Schirmer2010} it is crucial to have good analytic approximations of the open system dynamics available, along with a fitting physical intuition. Given that many real-world applications are embedded in complex environments with short correlation times, a Markovian treatment is here a natural point of departure. While numerical treatments yield quantitatively good results \cite{Kosloff2003,Vacchini2007,Eisfeld2009}, they become generally intractable as a large parameter space is to be explored or the system size increases. Moreover, numerical approaches alone are generally not suited to yield deeper insights into the general mechanisms at work. 

In this letter we develop a framework to obtain reliable analytic approximations to arbitrary Markovian master equations, based on a formal Dyson-like expansion of the quantum evolution. 
The key ingredient is an adaptive resummation of the series which optimizes its convergence even in the absence of a small parameter. The resulting decomposition into quantum jumps and periods of unperturbed evolution reflects the interplay of coherent and incoherent dynamics in a natural way and provides a physically meaningful interpretation. Based on the lowest order terms of the optimized expansion we obtain highly accurate approximations for the entire dynamics.

We start by presenting the general theory, and then illustrate its use and power by means of two nontrivial open quantum problems, reflection at a spatial detector \cite{Jacobs1999} and Landau-Zener tunneling with dephasing \cite{Kayanuma1984a,Kohler2007}. The resulting approximations are asymptotically exact and show errors on the per mil level over the entire range of parameters. This substantially improves existing approximations \cite{Jacobs2010,Jauslin2007}.

%**************************************************************************************

\emph{Jump expansion.---}The dynamics of a Markovian open quantum system can be described by a master equation $\dot \rho_t = \mathcal{L}(t)\rho_t$ with \cite{Breuer2002}
\be
	\mathcal{L}(t) \rho_t = -\frac i \hbar \left[ H(t), \rho_t \right] + \sum_{j=1}^N L_j \rho_t L_j^\dagger - \frac 1 2 \left\{ L_j^\dagger L_j, \rho_t\right\}. \label{eq:mastereq}
\ee
The operators $L_j$, which account for the incoherent influence of the environment, are not unique \footnote{We assume the $L_j$ to be time independent for notational simplicity; all derivations hold true for arbitrary $L_j(t)$.}. In particular, \eqref{eq:mastereq} is invariant under the transformation
\begin{align}
  L_j &\rightarrow L_{j,\boldsymbol \alpha} = L_j + \alpha_j\\
  H(t) &\rightarrow H_{\boldsymbol \alpha} (t) = H(t) - \frac{i\hbar}{2}\sum_{j=1}^N \left(\alpha_j^\ast L_j - \alpha_j L_j^\dag \right),
\label{eq:lotrafo}
\end{align}
with $\boldsymbol \alpha = (\alpha_1,\ldots,\alpha_{N}) \in \mathbbm{C}^N$. 

A formal expansion of the solution $\rho_t$ is obtained by decomposing the generator (\ref{eq:mastereq}) into any two parts, $\mathcal{L}(t) = \mathcal{L}_{\boldsymbol {\alpha}} (t) + \mathcal{J}_{\boldsymbol {\alpha}}$ (where $\boldsymbol {\alpha}$ labels different possible decompositions). Expanding the propagator $\mathcal T \exp [\int_0^t \mathcal L (t') \dd t']$ for a fixed decomposition yields a Dyson-like series, the \emph{jump expansion} \cite{holevo2001,Hornberger2009}                                                                                                                                                                                                                                                                                                                                             
\begin{align}
	\rho_t &= \sum_{n=0}^\infty \rho^{(n)}_t,\text{ with} \label{eq:jumpexpansion}\\
	\rho^{(n)}_t &= \int_{0}^{t} \dd t_n \, \mathcal{U}_{\boldsymbol {\alpha}} (t,t_n) \mathcal{J}_{\boldsymbol {\alpha}}\; \rho^{(n-1)}_{t_n} .
	\label{eq:jumprecursion}
\end{align}
Here ${\mathcal{U}_{\boldsymbol {\alpha}} (t_2,t_1) = \mathcal{T} \exp[ \int_{t_1}^{t_2} \mathcal{L}_{\boldsymbol {\alpha}} (t')\dd t']}$ propagates $\rho$ from $t_1$ to $t_2$, $\mathcal{T}$ denotes time ordering, and $\rho^{(0)}_t = \mathcal{U}_{\boldsymbol {\alpha}} (t,0) \rho_{0}$.

In analogy to the usual Dyson series for unitary dynamics, constituents in \eqref{eq:jumpexpansion} can be viewed as periods of unperturbed dynamics $\mathcal{U}_{\boldsymbol {\alpha}}$ interspersed with random perturbations, or jumps, $\mathcal{J}_{\boldsymbol {\alpha}}$. The superscript $(n)$ labeling different orders in the jump expansion hence denotes the number of jumps associated with the respective term.

A natural decomposition of $\mathcal L$ is given by
\begin{align}
  \mathcal J_{\boldsymbol \alpha} \rho_t  &= \sum_j L_{j,\boldsymbol \alpha} \rho_t L_{j,\boldsymbol \alpha}^\dagger \equiv \sum_j \mathcal J_{j,\boldsymbol \alpha}\, \rho_t
\label{eq:cpdeco1}
  \\
  \mathcal L_{\boldsymbol \alpha}(t) \rho_t &= - \frac i \hbar \left(H^\text{eff}_{\boldsymbol \alpha} (t) \rho_t - \rho_t {H^{\text{eff}}_{\boldsymbol \alpha}}^\dag (t)\right),
\label{eq:cpdeco2}
\end{align}
where $H^\text{eff}_{\boldsymbol \alpha}(t) = H_{\boldsymbol \alpha}(t) - i\hbar/2 \sum_j L_{j,\boldsymbol\alpha}^\dag L_{j,\boldsymbol\alpha}$. It has the special property that both the jumps $\mathcal{J}_{j,\boldsymbol \alpha}$ and the unperturbed propagators $\mathcal{U}_{\boldsymbol \alpha}$ are completely positive (norm-decreasing) superoperators. All orders $\rho^{(n)}_t$ of the jump expansion are then unnormalized density matrices with weights $w_n(t) = \text{Tr} \rho^{(n)}_t$ adding up to $\text{Tr} \rho_t = 1$. The $w_n(t)$ may therefore be viewed as probabilities to register $n$ jumps until $t$. Indeed, the jump expansion with \eqref{eq:cpdeco1}, \eqref{eq:cpdeco2} provides a physically meaningful measurement interpretation of the open system dynamics \cite{Hornberger2009}, which is essentially equivalent to the quantum trajectory approach \cite{carmichael1993,RevModPhys.70.101}.

%**************************************************************************************

\emph{Resummation.---}In contrast to usual implementations of the Dyson expansion, the decomposition of $\mathcal{L}(t)$ into $\mathcal{L}_{\boldsymbol \alpha} (t) + \mathcal{J}_{\boldsymbol \alpha}$ does not involve a small parameter. This renders the series~\eqref{eq:jumpexpansion} with its doubtful convergence properties of little practical use. However, we will see that a rapidly convergent series can be generated by choosing optimal decompositions $\boldsymbol{\tilde{\alpha}}$ of $\mathcal L(t)$. 

As discussed above, the different orders in~\eqref{eq:jumpexpansion} are characterized by weights $w_n(t)$, which satisfy $w_n(0) = \delta_{n,0}$. To maximize the contribution of the lowest order around $t=0$, the rate of change $-\partial_t w_0(t)|_{t=0}$ must therefore be minimal. This optimization condition determines $\boldsymbol{\tilde{\alpha}}$ at $t=0$. For $t>0$, $\boldsymbol {\tilde{\alpha}}$ is a function of both time and the jump record $\mathfrak R^n = (j_1,t_1;\ldots;j_n, t_n)$, which collects the sequence of past jumps $\mathcal J_{j,\boldsymbol\alpha}$ and their associated times. The weights of the $n$ lowest orders are maximized by minimizing $-\partial_t \sum_{m=0}^n w_m(t)$ at all times, a condition which will eventually yield the optimal $\boldsymbol {\tilde{\alpha}} (t,\mathfrak R^n)$.

Since the the jump operators will be conditioned on the record of past jumps, the operator decomposition underlying the expansion changes from term to term. It is not at all obvious why such an adaptive expansion should constitute a solution of the master equation. To see that this is indeed the case we write the order $\rho_t^{(n)}$ in terms of the record-conditioned branches $\rho_t^{(\mathfrak R^n)} $,
\begin{equation}
	\rho_t^{(n)} = \sum_{j_1, \ldots, j_n} \int_0^t \dd t_n \ldots \int_0^{t_2} \dd t_1 \; \rho_t^{(\mathfrak R^n)} \equiv \sum_{\{\mathfrak R^n\}}\rho_t^{(\mathfrak R^n)}, 
\nonumber
\end{equation}
where $\rho_t^{(\mathfrak R^n)} = \mathcal U_{\boldsymbol {\tilde \alpha} (t, \mathfrak R^n)} (t, t_n) \mathcal J_{j_n,\boldsymbol{\tilde\alpha}(t_n,\mathfrak R^{n-1})}\rho_{t_n}^{(\mathfrak R^{n-1})}$. The time derivative of $\rho_t^{(n)} $ decomposes into an $\mathfrak R^n$-  and an $\mathfrak R^{n-1}$-conditioned term
\be
  \partial_t \rho_t^{(n)} = \sum_{\{\mathfrak R^n\}}\mathcal L_{\boldsymbol{\tilde \alpha}(t,\mathfrak R^n)}(t) \rho_t^{(\mathfrak R^n)} +\!\!\!\!\!\sum_{j_n,\{\mathfrak R^{n-1}\}} \!\!\!\!\! \mathcal{J}_{j_n,\boldsymbol{\tilde \alpha}(t,\mathfrak R^{n-1})}\; \rho^{(\mathfrak R^{n-1})}_t.
\nonumber
\ee
Combining the  $\mathfrak R^n$-conditioned terms in $\partial_t \rho_t^{(n)}$ and in $\partial_t \rho_t^{(n+1)}$  yields $\sum_{\{\mathfrak R^n\}}\mathcal L(t) \rho_t^{(\mathfrak R^n)}$, which is independent of $\boldsymbol{\tilde \alpha}$. Taking the sum over all orders, we obtain the master equation $\partial_t \rho_t = \mathcal L(t) \rho_t$. 

To find the optimal adaptive decomposition $\boldsymbol{\tilde \alpha} (t,\mathfrak R^n)$, we now consider the rate of change
\begin{eqnarray}
  -\partial_t \sum_{m=0}^n w_m(t) 
  &=& \!\!\! \sum_{i,\{\mathfrak R^n\}}\! \text{Tr}\left\{ \left( L_i^\dag L_i + \left| \tilde \alpha_i(t,\mathfrak R^n)\right|^2\right)\rho_t^{(\mathfrak R^n)}\right\} \nonumber\\
  &&\!\!+ 2\,\text{Re}\left\{ {\tilde \alpha_i}^\ast (t,\mathfrak R^n)\text{Tr}\left( L_i \rho_t^{(\mathfrak R^n)}\right)\right\}.
\end{eqnarray}
The minimum is attained for the choice
\be
  \tilde \alpha_i (t,\mathfrak R^n) = \left. -\text{Tr}\left( L_i \rho_t^{(\mathfrak R^n)}\right) \middle/ \text{Tr} \rho_t^{(\mathfrak R^n)}\right. ,
  \label{eq:optalpha}
\ee
which ensures optimal convergence of the jump expansion up to $n^\text{th}$ oder.

We note that this type of jump operators can also be found in the context of stochastic unravelings, by requiring the jumps to map into orthogonal subspaces \cite{Diosi1986,Hornberger2010} or by minimizing the entropy production of an associated measurement \cite{PhysRevLett.74.4827}. Moreover, different update rules for the jump operators are obtained if one optimizes for different objectives, such as in feedback control \cite{Wiseman2005}. The crucial point for our purposes is that the adaptive decomposition defined by \eqref{eq:optalpha} gives rise to the most efficient resummation of the jump expansion in the sense that even a truncated series captures the essential part of the entire dynamics.

In addition to optimizing the convergence of the jump expansion, our goal is to make its solution analytically tractable. A trade-off of both aspects is obtained through an unbiased elimination of the state-dependence of $\boldsymbol{\tilde \alpha}$ in \eqref{eq:optalpha} by assuming complete ignorance, $\rho \propto \mathbbm 1$, except immediately after a jump. This way  \eqref{eq:optalpha} simplifies to
\be
  \alpha_i (\boldsymbol j^n) = -\frac{\text{Tr}\left( L^\dag_{j_n,\boldsymbol \alpha(\boldsymbol j^{n-1})} L_i L_{j_n,\boldsymbol \alpha (\boldsymbol j^{n-1})}\right)}{\text{Tr} \left( L^\dag_{j_n,\boldsymbol \alpha (\boldsymbol j^{n-1})} L_{j_n,\boldsymbol \alpha(\boldsymbol j^{n-1})}\right)}.\label{eq:bestalpha}
\ee
The operators $L_{i,\boldsymbol \alpha}$ are then piecewise constant between successive jumps and they depend only on the \emph{sequence} $\boldsymbol j^n = (j_1, \ldots, j_n)$ of previous jumps. Passing from \eqref{eq:optalpha} to \eqref{eq:bestalpha} thus amounts to using only the most relevant part of the information contained in the record $\mathfrak R^n$.

Other simplifications are conceivable, \eg incorporating different parts of the  information in $\mathfrak R^n$ for the adaptive resummation. However, we find that \eqref{eq:bestalpha} proves surprisingly powerful in a number of applications. In particular, already the first two orders give rise to highly accurate approximations for the spatial detection of particles and for Landau-Zener tunneling with dephasing.

%**************************************************************************************

\emph{Reflection by measurement.---}Our first application demonstrates two important aspects of the resummation: $(i)$ Its rapid convergence ensures that the first terms provide good approximations, and $(ii)$ it offers a physical picture of the underlying dynamics, which can serve as the starting point for an analytical treatment. Specifically, we find that the resummation replaces frequent but insignificant projective measurements of a particle on the left and right half space by rare but decisive transits across the measurement boundary.

Assuming that projective measurements onto the positive and negative half-axis occur with rate $\gamma$, and that the outcomes are discarded, the motion of a free particle can be described by a master equation with $H = \hat p^2 / 2 m$, $L_1 = \sqrt \gamma \Theta (-\hat x)$ and $L_2 = \sqrt \gamma \Theta (\hat x)$. Equation \eqref{eq:bestalpha} then implies that the optimized operators $L_{i,\boldsymbol \alpha(\boldsymbol j^n)}$ are identical, alternating between $\sqrt \gamma \Theta (\hat x)$ and $\sqrt \gamma \Theta (-\hat x)$ upon every jump.  As a consequence, the particle must traverse the measurement boundary $x=0$ between two jumps, with the number of transits given by the jump count. 

The particle now experiences only a \emph{finite} number of transits before going off to the left or to the right, as can be confirmed by a numerical unraveling. Compared with the unbounded number of jumps of the original expansion this reflects the rapid convergence of the resummation. Moreover, since every second term contributes to reflection (no transit, two transits, etc.), the reflection probability is the sum of the weights $P_{2n} = w_{2n}(t\to\infty)$ of all even jumps terms.

\begin{figure}[tb]
 \includegraphics[width=\columnwidth]{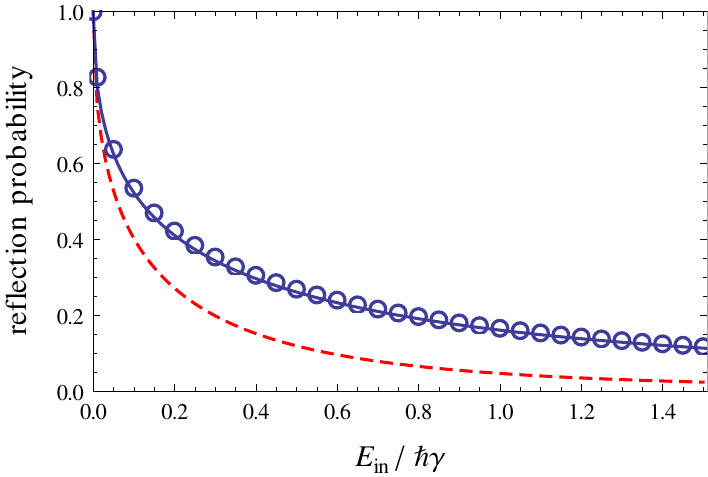}
  \caption{Reflection probability of a free particle with kinetic energy  $E_\text{in}$ 
  at the boundary of a projective left-right measurement with rate $\gamma$. Open dots: numerically exact values; solid line: prediction \eqref{eq:totprob} of the adaptive jump expansion; dashed line: leading order approximation (\ref{eq:imagreflec}) \cite{Jacobs2010}. The result \eqref{eq:totprob} is asymptotically exact, with a maximal deviation (at $E_\text{in}/\hbar\gamma \approx 0.2$) of below one percent.}
 \label{fig:reflection}
\end{figure}

As follows from \eqref{eq:cpdeco2}, the lowest order contribution is determined by the reflection at the imaginary potential step $-i\hbar \gamma \,\Theta(\hat x)$. The corresponding stationary solution $\ket{\psi^{(0)}_{k_0}}$, consisting of a plane wave to the left and an exponential tail to the right of the step, gives the reflection probability
\be
  P_0 (k_0) = \left| \frac{1-\sqrt{1 + 2 m \gamma i/\hbar k_0^2}}{1+\sqrt{1 + 2 m \gamma i/\hbar k_0^2}} \right|^2,
  \label{eq:imagreflec}
\ee
as a function of the incoming wave number $k_0$. The state after the first jump $\ket{\psi^{(1)}_{k_0}}$ is obtained by projecting $\ket{\psi^{(0)}_{k_0}}$ onto the right hand side:
\be
 \bra{x}\psi^{(1)}_{k_0}\rangle = \Theta(x) \mathcal N^{\frac 1 2} \exp\left( i x \sqrt{k_0^2+2m \gamma i/\hbar} \right),
 \label{eq:psi1x}
\ee
where $\mathcal N = 2 \,\text{Im}\sqrt{k_0^2 + 2m \gamma i/\hbar}$. Due to the adaptive update of the jump operators, the imaginary potential step then changes to $-i\hbar \gamma \,\Theta(-\hat x)$, which now acts as a scattering potential for $\ket {\psi^{(1)}_{k_0}}$. Since $\bra{k}\psi^{(1)}_{k_0}\rangle$ is a Lorenzian wave packet, one must treat the incoming ($k<0$) and outgoing ($k>0$) parts separately with factors $P_0(k)$ and 1. This yields
\begin{align}
\label{eq:imagtworeflec}
  P_{1} = &\bigg\{\int_{-\infty}^\infty \!\!\dd k \Big(\Theta(-k) P_0 (k) +\Theta(k)\Big)\left| \bra{k}\psi^{(1)}_{k_0}\rangle \right|^2 \\
  &+ \int_0^\infty \!\!\dd t \,2\gamma e^{-2\gamma t} \bra{\psi^{(1)}_{k_0}} \hat C(t) \ket{\psi^{(1)}_{k_0}} \bigg\}(1-P_0(k_0)).
\nonumber
\end{align}
The reflection probability of $\ket {\psi^{(1)}_{k_0}}$, given within curly brackets, involves a small correction term with $\hat C(t) = \Theta(-\hat k) \Theta(-\hat x - \hbar \hat k t/m)\Theta(\hat k) + \Theta(\hat k) \Theta(-\hat x - \hbar \hat k t/m)\Theta(-\hat k)$, accounting for the norm decay during the scattering process at the imaginary potential. All higher order terms depend on the specific jump times due to the non-stationarity of $\ket{\psi^{(1)}_{k_0}}$. However, the reflection probabilities of $\ket{\psi^{(n)}_{k_0}}$ can be approximated by that of  $\ket{\psi^{(1)}_{k_0}}$ in \eqref{eq:imagtworeflec}, such that we have $P_{n+1} \approx (1-P_1/(1-P_0(k_0))) P_n$ and hence the total reflection probability
\be
	\sum_n P_{2n} \approx 1 - \frac { \left(1-P_0(k_0)\right)^2}{2-2P_0(k_0)-P_1}.
	\label{eq:totprob}
\ee

In Fig.~\ref{fig:reflection} we compare this readily accessible approximation to the results of a time-consuming simulation of the master equation. We find that the error remains below 1\% over the entire parameter range---a considerable improvement of the known leading order approximation \cite{Jacobs2010}. It is remarkable, that this degree of accuracy is obtained with only the two lowest order reflection probabilities, highlighting the rapid convergence of the optimized jump expansion.

%**************************************************************************************

\emph{Landau-Zener transitions with dephasing.---}The second application is a bit more involved. It demonstrates that the adaptive resummation reformulates incoherent dynamics in a way that guides our intuition towards highly accurate analytic approximations.

The Landau-Zener problem involves a pair of states $\{\ket{1},\ket{2}\}$ with constant coupling $\alpha$ and time-dependent energy separation $v t$. Rescaling time as $ \tau = t \sqrt{v/\hbar}$ one finds that the Schr\" odinger equation
\be
\label{eq:LZresc}
  \frac{\partial}{\partial \tau} \ket{\psi} 
  = -i\left(\frac{\tau}{2} \sigma_z + \sqrt \delta \, \sigma_x\right) \ket{\psi} 
\ee
involves a single dimensionless quantity $\delta = \alpha^2 / \hbar v$, called adiabaticity parameter. Starting out in $\ket 1$ or $\ket 2$ in the remote past, the probability to experience a transition is $P(\delta,0) = 1 - \exp ( - 2 \pi \delta)$ as $\tau \to \infty$ \cite{MajoranaLZ,*ZenerLZ,*LandauLZ,*StueckelbergLZ}. 

The coherent evolution \eqref{eq:LZresc} will now be supplemented by \emph{dephasing}, as described by the operator $L = \sqrt{\gamma/2} \sigma_z$ (or equivalently by $L = \sqrt{2\gamma} \ket 2 \bra 2$). This models the influence of a continuous energy measurement, or of rapidly fluctuating energy levels with dephasing rate $\gamma$. As a result, the transition probability $P(\delta,\gamma)$ depends on $\delta$ and $\gamma$.

Before applying the jump expansion~\eqref{eq:jumpexpansion} and resummation \eqref{eq:bestalpha}, it is helpful to consider the strong dephasing limit. Representing $\rho$ as a Bloch vector, its $x$- and $y$-components can be adiabatically eliminated for $\gamma \to \infty$, by setting $\dot x=\dot y=0$. This leads to a closed evolution equation for the $z$-component describing the populations. Integration yields $P(\delta,\infty) = (1-\exp(-4\pi\delta))/2$, in agreement with the Landau-Zener transfer probability derived in the context of rapidly oscillating energy levels \cite{Kayanuma1984a}. It reflects the suppression of quantum tunneling by frequent measurements, known from the quantum Zeno effect. With the exact expressions for $P(\delta,0)$ and $P(\delta,\infty)$ at hand, we can now tackle the $\gamma$-dependence of the transfer probability.

Similar to the previous application, the adaptive resummation~\eqref{eq:bestalpha} causes the jump operator to alternate between $2\sqrt \gamma \ket 2 \bra 2$ and $2\sqrt \gamma \ket 1 \bra 1$ upon every jump. Therefore, a Landau-Zener transition necessarily separates two consecutive jumps, which limits the number of jumps overall and assigns the \emph{odd} jump terms to contribute to the transition probability, \ie $P(\delta,\gamma) =  \sum_n P_{2n+1}$, with $P_n = w_n(\tau \to\infty)$.

The incoherent dynamics can thus be viewed as an alternating sequence of Landau-Zener tunnelings and jumps. This insight allows us to map the problem to a classical inhomogeneous Markov process with two alternating, time-dependent rates $\lambda_0(\tau)$ and $\lambda_1(\tau)$, corresponding to the time dependent Landau-Zener transfer rate and to the jump rate, respectively, 
\begin{align}
 \partial_\tau p_{2n}(\tau) &= \lambda_1(\tau) p_{2n-1} (\tau) - \lambda_0(\tau) p_{2n}(\tau) \label{eq:inhmpev}\\
 \partial_\tau p_{2n+1}(\tau) &= \lambda_0(\tau) p_{2n} (\tau) - \lambda_1(\tau) p_{2n+1}(\tau), \label{eq:inhmpunev}
\end{align}
with $p_0(\tau) = \exp(-\Lambda_0(\tau))$, $\Lambda_i(\tau) = \int_0^\tau \lambda_i (t) \dd t$.

It is natural to assume that the Landau-Zener transitions occur only during a characteristic time interval of length $\tau^\ast$ when the energy levels are close. Therefore, the probability $P_n$ for $n$ \emph{jumps} until $\tau = \infty$ is equal to the probability for $n$ \emph{transitions} until $\tau = \tau^\ast$, \ie $P_n = p_{2n-1}(\tau^\ast)+p_{2n}(\tau^\ast)$. 

To recover the strong dephasing limit, take $\lambda_1(\tau) \gg \lambda_0(\tau)$. After adiabatically eliminating all $p_{2n+1}$ an inhomogeneous Poisson process with rate $\lambda_0(\tau)$ is thus obtained for the $P_n$. Comparing its probability distribution at time $\tau^\ast$, $P_n =\Lambda_0^n \exp(-\Lambda_0)/n!$ (where $\Lambda_i \equiv  \Lambda_i(\tau^\ast)$), with the previously derived $P(\delta,\infty)$ implies $\Lambda_0 = 2 \pi \delta$.

\begin{figure}[tb]
   \centering
   \includegraphics[width=\columnwidth]{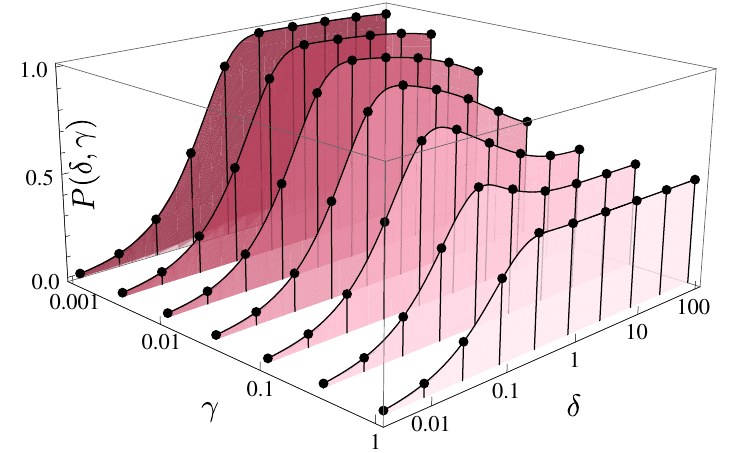}
   \caption{Landau-Zener tunneling probability $P$ as a function of dephasing rate $\gamma$ and adiabaticity parameter $\delta$. Numerically exact results (dots) are compared to the prediction $P = \sum_n P_{2n+1}$ from the jump expansion (black lines), with the approximation \eqref{eq:stochprobs} and parameters $\Lambda_0 = 2\pi \delta$, $\Lambda_1 = \gamma \tau^\ast$, and $\tau^\ast$ in \eqref{eq:tausj1}. The deviation does not exceed 0.5\% over the entire parameter range.}
   \label{fig:LZDP}
\end{figure}

By symmetry, the limit of strong Landau-Zener transitions, $\lambda_0(\tau) \gg \lambda_1(\tau)$, yields an inhomogeneous Poisson process with rate $\lambda_1(\tau)$. To obtain Poissonian asymptotics in both $\Lambda_0$ \emph{and} $\Lambda_1$ we require an exponential behavior of $P_n$ in $\Lambda_1$ as $\Lambda_1 \to \infty$,
 \be
 	P_n = \frac {\Lambda_0^n}{n!} e^{-\Lambda_0} \left(1 + f_n(\Lambda_1) e^{-\Lambda_1}\right).
 \ee
Here, $f_n(\Lambda_1)$ is a polynomial in $\Lambda_1$, with coefficients depending on $\Lambda_0$. The first $n-1$ of these coefficients are determined by the transition probability for $\gamma = 0$ and by the asymptotic behavior $P_n = \mathcal O (\Lambda_1^{n-1})$ as $\Lambda_1 \to 0$ of the non-decreasing Markovian (birth) process. By requiring that all higher orders of $f_n$ vanish we obtain
 \be
 	P_n = \frac{\Lambda_0^n e^{-\Lambda_0}}{n!}\!  \left[1\! -\! \frac {\Gamma(n,\Lambda_1)}{(n-1)!}\right]
 		 + \frac{\Lambda_1^{n-1}e^{-\Lambda_1}}{(n-1)!}\!  \left[1\! -\! \frac {\Gamma(n,\Lambda_0)}{(n-1)!}\right],
\label{eq:stochprobs}
 \ee
where $\Gamma(n,x)$ is the incomplete gamma function. This result shows the required asymptotic Poissonian behavior of $P_n$ in $\Lambda_0$ and $\Lambda_1$.

Identifying $\Lambda_1$ with $\gamma \tau^\ast$, the only free parameter is the characteristic Landau-Zener interaction time $\tau^\ast$. It can be determined by taking the derivative $\partial_\gamma P_1$ at $\gamma = 0$ in \eqref{eq:stochprobs}. $P_1$ is the weight of the first order term of the jump expansion, for which an exact analytic expression exists in terms of the parabolic cylinder functions \cite{akulin_tdsystems}. With the help of their asymptotic expansions \cite{abramowitz} we obtain
\be
	\tau^\ast \approx \frac{\pi \tanh(5/2 \delta)\sqrt{\delta\left(1-e^{-16 \delta}\right)}}{1-(1+2 \pi \delta) e^{-2 \pi \delta}}.
\label{eq:tausj1}
\ee

Fig.~\ref{fig:LZDP} compares the derived approximate transfer probability, as follows from \eqref{eq:stochprobs} with $\tau^\ast$ given by \eqref{eq:tausj1}, to the numerically exact value for $P(\delta,\gamma)$. One finds a striking agreement over the whole parameter range; the deviations remain below $0.5 \%$ and the asymptotic behaviors match exactly. This high quality of the approximation, which  substantially improves existing ones \cite{Jauslin2007}, makes it viable for practical applications such as quantum control tasks. Moreover, it suggests that the particular birth model \eqref{eq:inhmpev}, \eqref{eq:inhmpunev}, which was inspired by the adaptive resummation, captures the essence of the dynamics in this problem.

%**************************************************************************************

\emph{Conclusion.---}The general derivation and the worked out examples suggest that the adaptive expansion method provides the adequate way of unveiling the interplay of coherent and incoherent quantum dynamics. By guiding our intuition it serves as a natural starting point for efficient, analytically tractable approximation schemes, which may well advance the purposeful employment of incoherent processes.

\bibliographystyle{apsrev4-1}

\end{document}